\documentclass{aa}
\usepackage{graphicx}
\usepackage{xspace}
\usepackage{tikz}

\usepackage{txfonts}

\usepackage{natbib}
\usepackage{color}
\usepackage{fancyvrb}
\graphicspath{{./fig/}}
\def\pgas{p_{\rm gas}}
\def\pturb{p_{\rm turb}}
\def\ptot{p_{\rm tot}}

\newcommand{\muram}{MURaM\xspace}
\newcommand{\Mextend}{$\mathcal M_{\rm extend}$\xspace}

\newcommand{\Mpatch}{$\mathcal M_{\rm patch}$\xspace}
\newcommand{\muHz}{$\mu$Hz\xspace}

\begin{document} 

\title{
Estimating the nonstructural component of the helioseismic surface term using hydrodynamic simulations
}

\titlerunning{Surface term}

\author{J. Schou \and A. C. Birch}

\date{\today}

\institute{
Max-Planck-Institut f\"ur Sonnensystemforschung, Justus-von-Liebig-Weg~3, 37077 G\"ottingen, Germany\\
\email{schou@mps.mpg.de}
}

% \abstract{}{}{}{}{} 
% 5 {} token are mandatory
 
\abstract
{
As the amount of asteroseismic data available continues to grow, the inability to accurately model observed oscillation frequencies is becoming a critical problem for interpreting these frequencies. 
A major component of this problem is the modeling of the near-surface layers.
}
{
Our aim is to develop a method to estimate the effect of the near-surface layers on oscillation frequencies.
}
{
In the proposed method we numerically estimate eigenfunctions in 3D hydrodynamic simulations.
We match those to the eigenfunctions calculated from the classic equations applied to the horizontal averages of the structure variables.
We use this procedure to calculate the frequency perturbation resulting from the dynamical part of the interaction of the oscillations with near-surface convection.
As the last step we scale the numbers to the Sun.
To provide a qualitative test of our method we performed a series of simulations, calculated the perturbations using our procedure, and compared them to previously reported residuals relative to solar models.
}
{
We find that we can largely reproduce the observed frequency residuals without resorting to poorly justified theoretical models.
We find that, while the calculations of Houdek et al. (2017, MNRAS, 464, L124) produce similar frequency perturbations, the density-pressure phase differences computed here do not match those of that work.
}
{}


\maketitle

%________________________________________________________________

\section{Introduction}

Soon after frequencies of solar oscillations became available, attempts were made 
to infer the internal structure of the Sun, typically using inverse methods. 
It was determined that the frequencies could not be fitted, and it was quickly realized that a major source of the problem was close to the surface \citep[for details see e.g.,][]{1997MNRAS.284..527C}.
It was not a surprise that the surface caused problems as the physics in the surface layers is complicated by nonadiabatic effects, high velocities in the granulation, and a rapid variation of the thermodynamic quantities
\citep{1984Sci...226..687B}.
Fortunately, as the effect is near the surface, it must have a particular functional form. Given this, a term with a suitably chosen parametrization was added to the expression for the mode frequencies used in the inversions and the parameters were fit as part of the inversion procedure. This term became known as the "surface term."
While this is clearly unsatisfactory from the point of view of a physical understanding, the number of measured frequencies is large enough that the loss of degrees of freedom could be tolerated.
As no data were available for other stars, where the loss of degrees of freedom would be more significant, a solution to the problem was not urgent and the approach allowed for significant advances to be made in inferring the deeper solar structure
\citep[see, for example][]{2005ApJ...621L..85B}.

\cite{1999A&A...351..689R} suggested that part of the problem could be solved by patching averages of 3D~hydrodynamic simulations onto 1D~evolutionary models.
Unfortunately, as they also pointed out, determining how this averaging should be performed is not trivial and the patching approach does not solve the entire problem.
In particular, the mode physics is not addressed. A particular problem they point out is that the wave perturbation to the turbulent pressure needs to be modeled.
As another example of a dynamical effect, \cite{2012ApJ...760L...1B} find that the small-scale turbulence causes an apparent net upflow near the solar surface and that this may explain a number of center-to-limb problems observed in helioseismology.

As oscillations started to be observed on other stars, the problem of the surface term became acute. In asteroseismology, few modes are observed and simply fitting a parametrized function results in the loss of a substantial fraction of the available information. Current approaches include scaling the solar surface term and  using parametrizations with a small number of terms \citep[e.g.,][]{2008ApJ...683L.175K,2014A&A...568A.123B}.

Another approach,  employed by
\cite{2014MNRAS.437..164P},
\cite{2015A&A...583A.112S}, \cite{2016A&A...592A.159B}, and
\cite{2017A&A...600A..31S}, for example, is to use the suggestion of \cite{1999A&A...351..689R} and to patch averaged 3D simulations onto 1D evolutionary models. Indeed, this has resulted in a significant improvement in the fits with residuals that were reduced from up to 12~\muHz below 4000~\muHz to about 3~\muHz. While this is an important advancement, the residuals are still orders of magnitude larger than the observed errors. 
Unfortunately, the various dynamical effects are still not properly modeled.
\cite{2017MNRAS.464L.124H} addresses
this problem based on a model of convection.\ However, while their approach does reduce the residuals down to about 2~\muHz, there are still significant residuals, and it is unclear how good the model is and if it is applicable to other stars.
Refinements of these approaches, such as including evolutionary effects, were also considered by \cite{2018MNRAS.481L..35J}, \cite{2019MNRAS.488.3463J}, and \cite{2020MNRAS.491.1160M}, among others.

In the method proposed in this paper we take the approach of \cite{1999A&A...351..689R} one step further by effectively patching the eigenfunctions from a 3D~simulation onto those from a 1D~model.
More precisely, we horizontally average 3D~models, as suggested by \cite{2017MNRAS.464L.124H} and \cite{2016A&A...592A.159B}, for example, and calculate the resonance frequencies as they did, compare those to the frequencies of the modes seen in the simulations, and scale the differences to the solar case,
thereby including all of the physical effects that were neglected in the averaging.
Doing so allows us to separate the issues pertaining to structural changes and inversions from the mode physics issues; the latter is the subject of interest of this paper.

In Section~\ref{sec:method} we discuss how to determine the frequency perturbations and apply them to the solar case.
In Section~\ref{sec:results} various results are presented, in Section~\ref{sec:disc} we discuss the results, and in Section~\ref{sec:conc} we conclude. 

Apart from the discussion in Section~\ref{sec:houdek-comp}, we do not discuss the behavior of the eigenfunctions near the surface and the physical interpretation of those results. We choose to defer this work to a later paper.

\section{Method}
\label{sec:method}

The method we propose consists of several steps, which are outlined in the subsections below.
In Section~\ref{sec:models} we describe the simulations used, in Section~\ref{sec:svd} how the eigenfunctions and frequencies are extracted is explained, in Section~\ref{sec:match} we discuss how to calculate the frequency perturbation relative to the horizontally averaged models using the standard equations, and in Section~\ref{sec:sun} how to scale the results to the solar case is explored.

\subsection{Models used}
\label{sec:models}

The MURaM simulations used here were computed as described in \cite{2013A&A...558A..48B} and   \cite{2017ApJ...834...10R} for a nonmagnetic Sun. However, they were modified in terms of spatial extent and resolution, as described below.

The simulations were performed in Cartesian boxes with periodic boundary conditions in the horizontal directions. The upper boundary is nonpenetrative, while the flows were allowed to pass through the lower boundary; the entropy of the upflows were selected so as to obtain the desired luminosity. Near the bottom, there is also a diffusive layer. The vertical grid is uniformly spaced.  We used a constant gravity of $27400$~cm~s$^{-1}$. No corrections were made for curvature effects.
For details on the numerical schemes, the treatment of the radiative transfer, equation of state, etc., we refer to the papers cited above.
 
Ideally, the simulations would be deep, as this would allow for a large number of radial orders.
Unfortunately, while it is not important for the present use to have an accurate stratification deep in the box, it is nonetheless best to avoid constraining the convection by having a box that is too narrow; the boxes should be at least twice, or preferably three times, as wide as their extent below the surface. Using deep boxes also implies that the maximum sound speed is larger, requiring smaller time steps, and this also increases the thermal relaxation time. 
Given these problems and the need for very long simulation times in order to resolve the modes and improve the signal-to-noise ratio (S/N), we thus had to compromise and instead chose to use several simulations with the same width, but different depths, in order to obtain a reasonable frequency coverage.
Table~\ref{table:models}
 summarizes the sizes, resolutions, time span, and other properties of the simulations, as well as the parameters used in the analysis.
 The main simulations used here are cases 10, 12, and 13. These are all very close in temperature and only slightly warmer than the Sun. Cases 11 and 15 are higher and lower resolution versions of case~10, intended to show that an adequate resolution was used.
 In order to save space, the snapshots were generally deleted and only averaged quantities were kept.
 
 Throughout this paper, we use the symbols $\pgas$ for the gas pressure, $\pturb$ for the turbulent pressure, $\ptot=\pgas+\pturb$ for the total pressure, $\rho$ for the density, $v_z$ for the vertical velocity, and $\Gamma_1$ for the first adiabatic exponent.
 To define a height coordinate that is consistent with the 1D models, where $\tau$ is not given, the height $z=0$ is given as the location where horizontally averaged density is $2.3 \times 10^{-7}$g cm$^{-3}$. This was chosen so as to make the average height of the $\tau_{\rm Rosseland}$ surface coincide with $z=0$ for the regular resolution cases, and to be consistent with \cite{2013A&A...558A..48B}.
 The distance from the center of the Sun is given by $r=z+R_\odot$, where $R_\odot$ is the solar radius.

\begin{table}
\caption{List of simulations used. The quantity $\Delta t_{\rm save}$ indicates the average spacing between the saved (analyzed) time steps.
As simulations that are started from a horizontally uniform state take time to develop convection and to reach a statistically steady state, a section at the beginning of each simulation was discarded based on a visual inspection.
These sections are not included in the length shown in the table.
We note that $\rm{T_{eff}}$ was calculated from the average of the energy flux over the last 1000 points of the simulation.
}
\label{table:models}      
\centering                          
\begin{tabular}{|l | r r r r r|}        
\hline                 
Case & 10 & 11 & 12 & 13 & 15 \\    
\hline                        
   Top (Mm) & 1.9 & 1.9 & 1.9 & 1.9 & 2.0 \\
   Bottom (Mm) & -4.1 & -4.1 & -6.1 & -5.0 & -4.0 \\
   Horiz. size (Mm) & 12 & 12 & 12 & 12 & 12 \\
   Vert. res (km) & 20 & 10 & 20 & 20 & 40 \\
   Horiz. res (km) & 60 & 30 & 60 & 60 & 120 \\
   Time (Ms) & 3.59 & 0.70 & 2.57 & 2.94 & 3.62 \\
   $\Delta t_{\rm save}$ (s) & 71.8 & 34.3 & 31.8 & 34.0 & 29.0 \\
   $z_{\rm extend}$ (Mm) & -3.5 & -3.5 & -5.5 & -4.5 & -3.5 \\
   $z_{\rm fit,min}$ (Mm) & -3.0 & -3.0 & -5.0 & -4.0 & -3.0 \\
   $z_{\rm fit,max}$ (Mm) & -2.0 & -2.0 & -4.0 & -3.0 & -2.0 \\
   $\rm{T_{eff}}$ (K) & 5867 & 5859 & 5880 & 5871 & 5835\\
\hline                                   
\end{tabular}
\end{table}

\subsection{Extraction of eigenfunctions}
\label{sec:svd}

One way to characterize the acoustic oscillations in the simulations is to measure their eigenfunctions and eigenfrequencies. 
In order to extract the eigenfunctions from a simulation, we compute the unweighted horizontal averages of $\pgas$, $\rho$, $\Gamma_1$, and $v_z$ at each saved time step and height in the simulation domain.   
At each time and height, we subtract the corresponding horizontal average to obtain a set of residuals from which the horizontally averaged $\pturb$ is computed. From that, we compute the horizontally averaged total pressure $\ptot=\pgas+\pturb$.

These horizontally-averaged quantities are averaged in time to obtain the mean thermodynamic quantities.\ We denote the resulting quantities with an overbar, for example $\overline{p}_{\rm gas}$ is the horizontally- and time-averaged gas pressure.

The residuals with respect to the time averages
are due to both the radial acoustic oscillations and the time-dependent convection. As described later in this section, we use frequency filtering to mostly separate these two contributions.  

\begin{figure}
\begin{tikzpicture}[x=0.11\columnwidth,y=0.11\columnwidth]
\node at (0, 0) {\includegraphics[width=0.95\columnwidth]{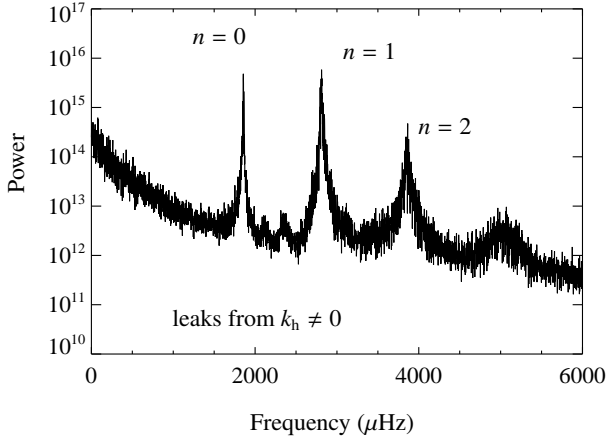}};
\node at (-1, 2.2) {$n=0$};
\node at (1, 2) {$n=1$};
\node at (2, 1) {$n=2$};
\node at (-0.5, -1.6) {leaks from $k_{\rm h} \neq 0$};
\node at (0.5, -3) {Frequency ($\mu$Hz)};
\node[rotate=90] at (-3.7,0.5) {Power};
\end{tikzpicture} 
\caption[]{
Power spectrum of the pressure perturbation for case~10, averaged over the depth range used for the SVD.
The power was smoothed to reduce the effect of the stochastic excitation.
The large peaks at 1856~\muHz, 2812~\muHz, and 3858~\muHz represent the modes with $n$=0, 1, and 2, respectively, where $n$ is the number of nodes in the displacement eigenfunction. The $n=3$ peak around 5000~\muHz is not analyzed further, as it is very close to the acoustic cutoff frequency.
The small peaks around 2100~\muHz and 2300~\muHz are leaks from modes with $k_{\rm h} \neq 0$.
\label{fig:power}}
\end{figure}

The wave component for each quantity is then linearly interpolated at each height to a uniform grid in time with twice the average resolution of the simulation grid and Fourier transformed.
Figure~\ref{fig:power} shows an example of a resulting power spectrum.
The power spectrum shows several resonances; these are the radial oscillations.  
The power spectrum also shows a number of weak peaks in between the main modes. An examination of the spectra of the modes with a horizontal wavenumber of $k_{\rm h} \neq 0$ shows that these peaks are due to leaks from modes with $k_{\rm h}=1$ and $k_{\rm h} =\sqrt{2}$ (with $k_{\rm h}$ in units of wavelengths across the box width), but they appear wider in the $k_{\rm h} =0$ spectra.
A likely explanation for their appearance is that the convection has large components at those horizontal wavenumbers, which interact with the corresponding modes, leading to power with $k_{\rm h} =0$. As the convection is time variable, it also leads to widening of the peaks.

Based on the stochastic nature of the modes, we expect each physical quantity to be separable in time (or equivalently frequency) and height near a resonance, with the former showing a Lorentzian power profile in frequency.  For example, in the neighborhood  of the $n^{\rm th}$~resonance, the Fourier transform of a physical variable $q$ has the form
\begin{equation}
q(z,\omega) \approx q_n^\prime(z) f_n(\omega) \; ,
\label{eq:sep}
\end{equation}
where $q_n^\prime$ is the eigenfunction (denoted here by a superscript $^\prime$) in height,
$f_n$ is the Fourier transform of the time dependence, and $\omega$ is the frequency. In general both $q_n^\prime$ and $f_n$ are complex.
The singular value decomposition (SVD) decomposes a signal into a sum of such terms:
\begin{equation}
q(z,\omega)=\sum_i w_i q_i^\prime(z) f_i(\omega) \; ,
\label{eq.svd}
\end{equation}
where the sum is taken over the singular vectors $q^\prime$ and $f$, which are generally complex, and the singular values $w$, which are real and sorted in decreasing value. In particular, the first term represents the best fit of Eq.~\ref{eq:sep}.
Performing the SVD on the entire dataset turns out not to be desirable for a number of reasons. The convective signal is very large near the photosphere and would introduce noise in the frequency dependence. There is significant convective power at low frequencies, which is not interesting for the purposes of this work. Also, as the eigenfunctions may be similar over a limited fitting range, they are not cleanly separated.   A joint SVD using all physical quantities would be another option, but this approach is not effective as the S/Ns for different physical quantities can be very different.

To overcome these issues, we use the pressure perturbation over the interior part of the simulations (excluding the lower 10\% and from 0.5~Mm below the surface and above) for the SVD, as this height range has a good S/N. Also, we only use a small range in frequency around each observed peak, which is $\pm$ four half width at half maximum (HWHM), but in no case is it smaller than $\pm 20$~\muHz or larger than $\pm 200$~\muHz. In all cases, it was found that a single component, by far, dominates the SVD, capturing above 99\% of the variance for modes below $4500$~\muHz and in some cases over 99.999\%. We thus only study the dominant component and suppress the subscript~$i$ in the following.
In other words, we simply use the SVD as a way to fit the data to the model given by Eq. \ref{eq:sep}.
As this procedure uses almost all of the power in a peak, it gives substantially less noisy results than using the spatial dependency at the peak frequency, for instance.
The complex conjugate of the dominant frequency dependence vector ($f$ in Eq.~\ref{eq.svd}) is then multiplied on all variables at all heights to obtain the vertical eigenfunctions for these variables.

Given the stochastic nature of the excitation of the modes, the mode frequency $\omega_n$ was then computed by fitting a Lorentzian to $
|f(\omega)|^2$, using a maximum likelihood fit \citep{1990ApJ...364..699A}.
Given the mode frequency, we also calculate the displacement eigenfunction $\xi = -v_z^\prime/(i\omega_n)$.

The SVD singular vectors are degenerate up to a phase factor.  For convenience, we shift the phases at each radial order to make the displacement eigenfunction real 50~gridpoints above the bottom.

\begin{figure}
\begin{center}
\begin{tikzpicture}[x=0.11\columnwidth,y=0.11\columnwidth]
\node at (0, 0) {\includegraphics[width=0.95\columnwidth]{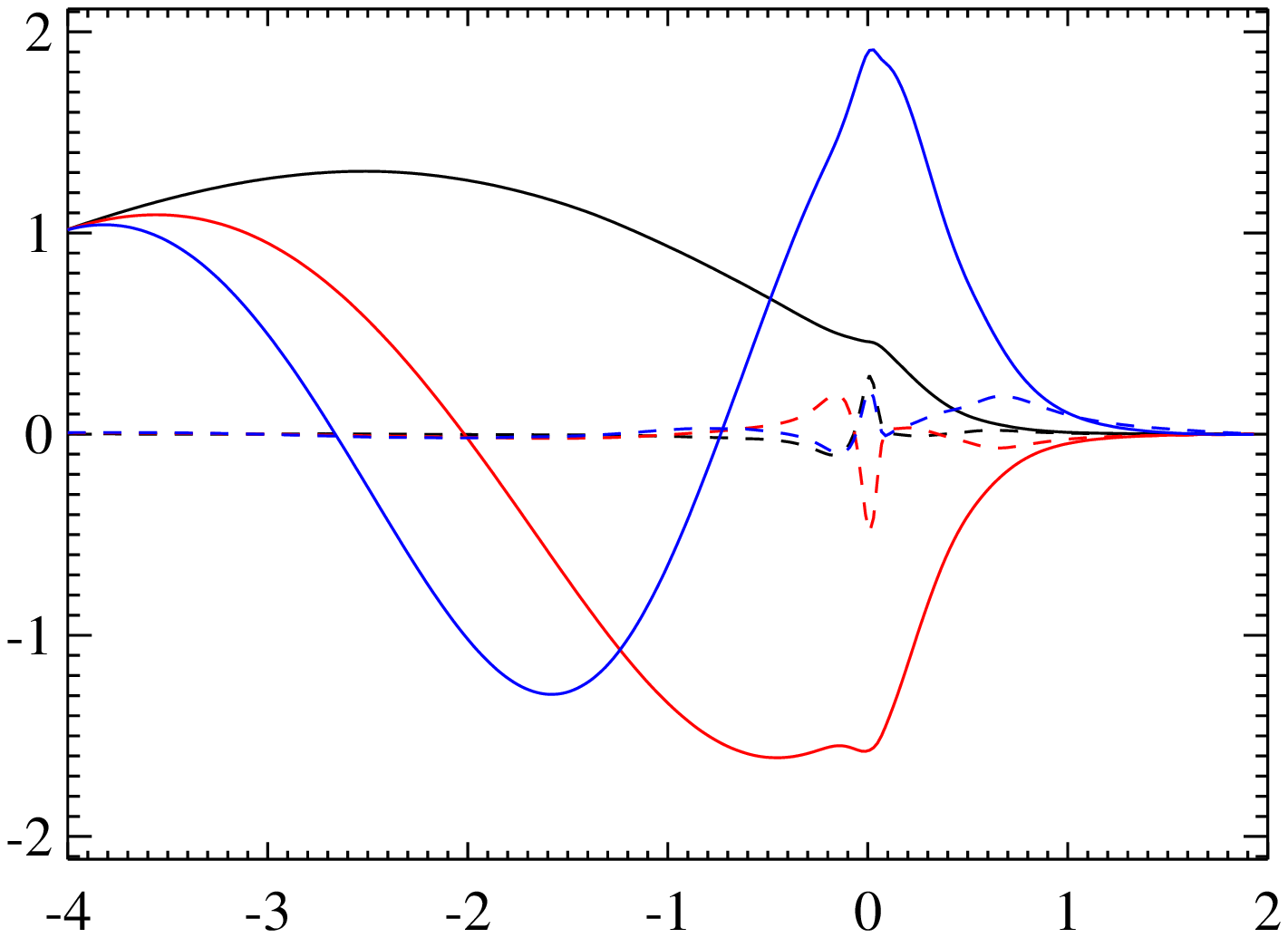}};
\node at (-1, 2.2) {$n=0$};
\node at (2.5, -1) {\textcolor{red}{$n=1$}};
\node at (2.5, 2) {\textcolor{blue}{$n=2$}};
\node at (0.5, -3) {Height (Mm)};
\node[rotate=90] at (-3.7,0.3) {$\sqrt{\bar{\rho}}\xi$};
\end{tikzpicture} 
\end{center}\caption[]{
Displacement eigenfunctions corresponding to the three largest peaks in Fig.~\ref{fig:power}.
The eigenfunctions were multiplied by the square root of the background density to improve the visibility and as the square of this quantity is proportional to the energy density.
Colors indicate the different modes.
Solid lines indicate the real part, and dashed lines illustrate the imaginary part.
For ease of display, the scaled eigenfunctions were normalized to unity at the bottom boundary.
\label{fig:ef}}
\end{figure}

Figure~\ref{fig:ef} shows examples of eigenfunctions. Beyond the classic behavior, which is defined by Eqs. \ref{eq.dxidz} through \ref{eq.csquared} below, the most notable feature is probably the large imaginary component near the surface, 
which was previously ascribed by \cite{1992MNRAS.255..603B} to nonadiabatic effects and nonlocal convection effects
and later by
\cite{2012ApJ...760L...1B} to the large horizontally-averaged vertical flow near the surface.
\cite{2001ApJ...546..576N} and \cite{2001ApJ...546..585S} also extracted eigenfunctions from simulations, but unfortunately they did not show the imaginary component.  The former paper also discusses some of the formalism needed to analyze the waves.
While this is clearly interesting in terms of understanding the relevant physics, we defer the study of the eigenfunctions to a future paper, except for a brief discussion that appears in Sect.~\ref{sec:houdek-comp}.

\subsection{Determination of the frequency perturbation}
\label{sec:match}

Our goal is to compute the frequency perturbation caused by the physics not captured by the simple patched models in a solar model that extends from the center of the star to the surface.
We tackle this calculation in two steps.  
In the first step,
we compute the change in the resonance frequencies caused by near-surface convection for an extended \muram model.  
In the second step, which is described in the next subsection,
we use a mode-mass correction to compute the frequency changes that would be expected for a solar model that extends down to the center of the star.
An advantage of this approach is that it avoids most issues resulting from different physics in the 3D box model and the 1D evolutionary models. In particular, issues relating to the curvature and variable gravity are avoided to lowest order, as are issues related to large scale structural problems.
Alternatives to this approach are discussed in Sec. \ref{sec:disc} and tests of various assumptions appear in the Appendix.

The background model, as well as the eigenfunctions computed in the previous section, have artifacts near the bottom boundary.
These artifacts are due to the effect of the boundary on the stratification and to the diffusive layer near the bottom.  If a layer at the bottom is simply removed, the boundary condition is ill defined.  
To mitigate these problems, and as the $n=0$ mode in any case does not have a zero crossing of the displacement eigenfunction within the simulation domain, we extend the \muram model downward to -50~Mm to capture at least one zero crossing of the displacement eigenfunction for each mode. This extended model, which is denoted with \Mextend, is constructed as follows:
First the part of the model below $z_{\rm extend}$ is removed. In all cases, $z_{\rm extend}$ is chosen to be 0.5 to 0.6~Mm above the bottom of the simulation, which ensures that the diffusive layer is not included and that the stratification at $z_{\rm extend}$ is very close to adiabatic at the averaged $\Gamma_1$. 
We then extrapolate $\Gamma_1$ linearly downward to a value of 1.6 at -9~Mm and keep it at that value below -9~Mm to correspond roughly to the results of deeper simulations. For the extension, we use a grid with the same vertical spacing as in the simulation. We note that the details of the extension have a negligible effect on the final result, as is demonstrated in Appendix \ref{app:tests}.
Assuming hydrostatic equilibrium, constant gravity, and perfect adiabatic stratification, we can extend the model downward to obtain $\ptot$ and $\rho$.
From that, we calculate
\begin{equation}
\frac{\overline{p}_{\rm turb}(z)}{\overline{p}_{\rm tot}(z)} = \exp\left [\frac{z-z_{\rm extend}}{3.03~{\rm Mm}}\right] \frac{\overline{p}_{\rm turb}(z_{\rm extend})}{\overline{p}_{\rm tot}(z_{\rm extend})} \; ,
\end{equation}
again to roughly match the behavior of the deepest simulation.
Finally we set $\overline{p}_{\rm gas}=\overline{p}_{\rm tot}-\overline{p}_{\rm turb}$.
For convenience, the averaged quantities are indicated by $\overline{\phantom{x}}$, even over the extended region, to make it clear as to which quantities are averaged and which are derived.

Given that the turbulent pressure is important for the structure, it is important to ask which $\Gamma_1$ should be used in the oscillation equations near the surface.
\cite{1999A&A...351..689R} considered two models.
The gas-gamma model (GGM) assumes that the turbulent and gas pressures behave similarly, that is that the effective $\Gamma_1$ is
\begin{equation}
\label{eq:CGM}
{\Gamma}_1^{\rm GGM} = \overline{\Gamma}_1\; ,
\end{equation}
where $\overline{\Gamma}_1$ is the horizontally averaged $\Gamma_1$. This is the approach taken by \cite{2016A&A...592A.159B}.

The reduced-gamma model (RGM) assumes that 
\begin{equation}
{\Gamma}_1^{\rm RGM} = \frac{\overline{p}_{\rm gas}}
{\overline{p}_{\rm tot}}
\overline{\Gamma}_1\;,
\end{equation}
in other words the turbulent pressure is not perturbed by the oscillations.
This approach is one of the cases investigated by \cite{2017MNRAS.464L.124H}, who also consider more elaborate models based on theories of convection.  As we have the eigenfunctions for all of the variables, we are able to address the accuracy of these approximations, as we briefly discuss in Section~\ref{sec:houdek-comp}.  In the equations below, we refer to a generic $\Gamma_1$. In the subsequent calculations, we consider both choices and compare each to the respective results shown in \cite{2016A&A...592A.159B} and \cite{2017MNRAS.464L.124H}. 

To calculate the computed eigenfunctions in the extended \muram model \Mextend,
we solve the standard ("classical") oscillation equations for radial ($k_h=0$) oscillations:
\begin{equation}
\frac{d\xi}{dz} = -\left (\frac{2}{r_{\rm c}}-\frac{1}{{\Gamma}_1 H_p} \right) \xi - \frac{1}{\overline{\rho} c^2} p^\prime
\label{eq.dxidz}
 \end{equation} and
\begin{equation}
\frac{dp^\prime}{dz} = \overline{\rho}(\omega^2-N^2+\omega^2_\Phi)\xi - \frac{1}{{\Gamma}_1 H_p} p^\prime,
\label{eq.dpdz}
\end{equation}
where
\begin{equation}
H_p^{-1} = -\frac{d ln \overline{p}}{dz}
\end{equation}
is the inverse of the pressure scale height,
and\begin{equation}
N^2=g\left(\frac{1}{{\Gamma}_1 \overline{p}} \frac{d\overline{p}}{dz} - \frac{1}{\overline{\rho}}\frac{d\overline{\rho}}{dz}\right)
\end{equation}
is the square of the buoyancy frequency, $r_{\rm c}$ is the local radius of curvature,
\begin{equation}
c^2 = {\Gamma}_1 \frac{\overline{p}}{\overline{\rho}}
\label{eq.csquared}
\end{equation}
is the squared sound speed, and we use $\overline{p}$ for the total pressure.
As the equations can be written in different forms by substituting variables, it is important to use a minimal number of variables, such that the results are insensitive to the rewrites. We therefore compute $N^2$ and $c^2$ from the mean pressure, $\Gamma_1$, and density, rather than using their averages from the simulations.
The quantity $\omega^2_\Phi$ is due to the perturbation to the gravitational potential; it is zero for the Cartesian box case (the acceleration due to gravity is constant in this case) and is given by $4\pi G \overline{\rho}$ for spherical models. The symbol $G$ denotes the gravitational constant.  In equation~(\ref{eq.dxidz}), the radius of curvature $r_c$ is infinity for the Cartesian box case and $r_c=r$ for the spherical case.

To extend the eigenfunctions of the \muram simulation down through the extended model \Mextend,
we solve equations~(\ref{eq.dxidz}) and~(\ref{eq.dpdz}) for the frequencies of each of the modes observed in the simulation.
As no boundary conditions are imposed, this yields two solutions.
We then compute the linear combination of these two solutions that best fits, in the least squares sense, the real part of the observed displacement eigenfunction over the fitting range given by $z_{\rm fit,min}$ and $z_{\rm fit,max}$ in Table~\ref{table:models}.
In other words, we effectively use the observed eigenfunction as an upper boundary condition.

\begin{figure}
\begin{center}
\begin{tikzpicture}[x=0.11\columnwidth,y=0.11\columnwidth]
\node at (0, 0) {\includegraphics[width=0.95\columnwidth]{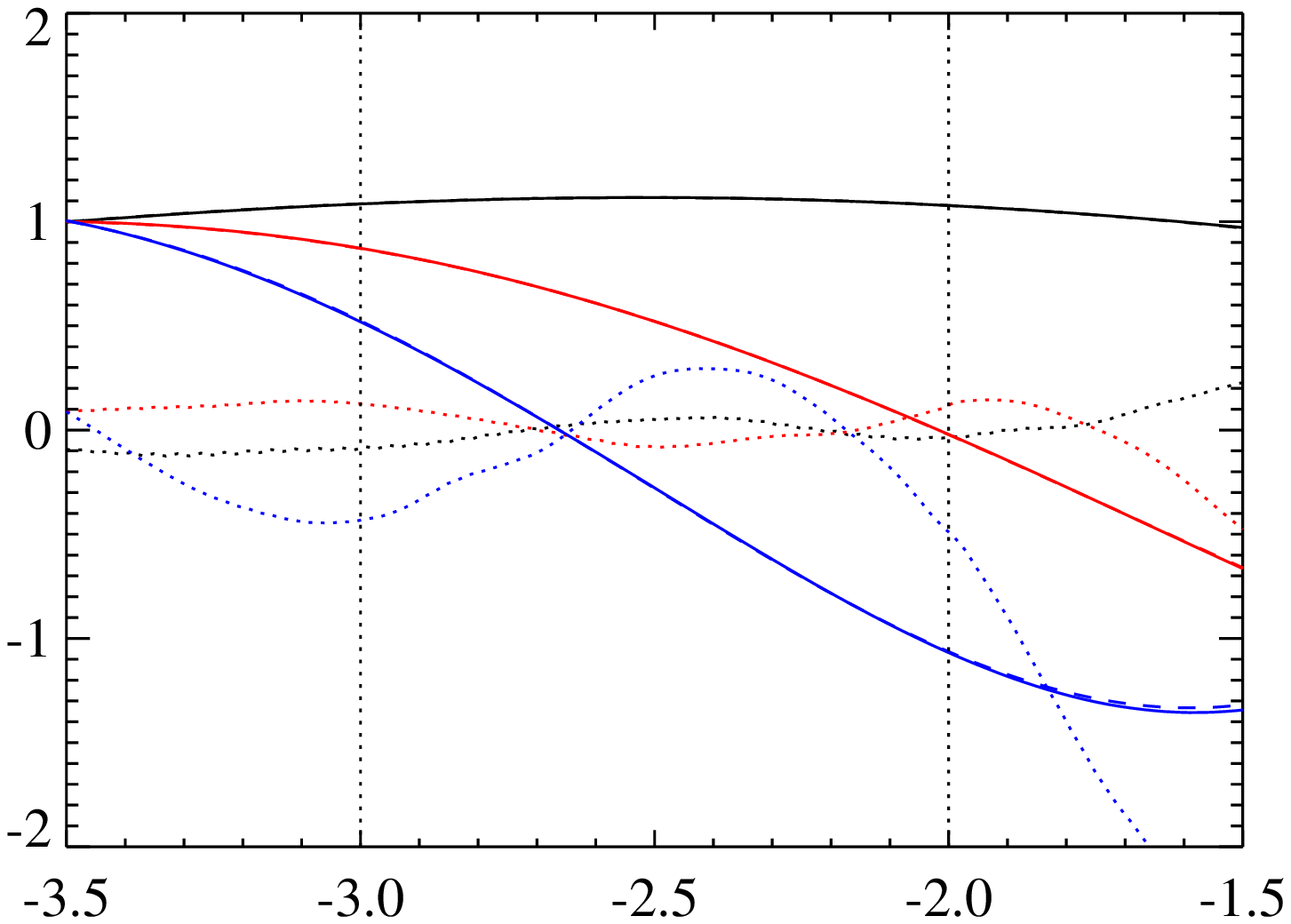}};
\node at (0, 2.0) {$n=0$};
\node at (3.2, -0.7) {\textcolor{red}{$n=1$}};
\node at (0.8, -0.7) {\textcolor{blue}{$n=2$}};
\node at (1, -3) {Height (Mm)};
\node[rotate=90] at (-3.7,0.3) {$\sqrt{\bar{\rho}}\xi$};
\end{tikzpicture} 
\end{center}\caption[]{
Examples of the real parts of the numerical eigenfunctions (solid) and fitted eigenfunctions (dashed, barely visible) for case~10 using the GGM approximation. Dotted curves show 100 times the difference. Vertical dotted lines show the fitting range.
\label{fig:efit}}
\end{figure}

Figure~\ref{fig:efit} shows examples of the fitted eigenfunctions computed using the above procedure. As can be seen, the fits are very good over the fitting range.

We compute the mode frequencies for the \Mextend stratification, neglecting the dynamical effects of the convection,
by solving the eigenvalue problem of equations~(\ref{eq.dxidz}) and~(\ref{eq.dpdz}).
As the bottom boundary condition, we use a zero in the displacement eigenfunction
at the location of a, deeply located, zero crossing in the fitted eigenfunction.
At the top, we match the eigenfunction to the solution for an isothermal atmosphere.  The matching is carried out at 0.5 Mm above the photosphere to roughly match the boundary conditions used by the results we compare to.
For the sake of comparison, we chose this boundary condition rather than a boundary condition that is higher in the atmosphere. 

As the mean stratification is the same in the two cases, the difference between a mode frequency in the \muram simulation and in \Mextend
is the frequency perturbation caused by the impact of the
effects that are not taken into account when averaging the background state and patching it onto a 1D model.  We denote this frequency shift as $\delta\omega_{\rm extend}$.

The mode masses in \Mextend are:
\begin{equation}
M_{\rm extend} = \int \overline{\rho}(z) \xi^2(z) d z 
,\end{equation}
where the (real-valued) eigenfunction $\xi$ is normalized to be one at the surface and the integral runs from the chosen zero crossing to the top of the box.  In the next section, we show that these mode masses are key to scaling the frequency shifts $\delta\omega_{\rm extend}$ to the solar case.

\subsection{Scaling to the Sun}
\label{sec:sun}

The frequency perturbation $\delta\omega_{\rm extend}$ is, however, not the one expected for the Sun as the model is truncated in depth.  We use a simple mode-mass scaling to correct for the truncation.   To see the intuitive motivation for our approach, consider the eigenvalue problem for radial modes $F\xi = \omega^2\xi$ where $F$ is the wave equation operator and $\omega$ is the normal-mode frequency.  The first-order perturbation theory relationship between a perturbation to the wave-equation operator $\delta F$ and the resulting perturbation to the square of the mode frequency $\delta\omega^2$ is
\begin{equation}
\delta\omega^2 = 
\frac{\int \xi \delta F(\xi) \rho r^2 dr}{M_{\rm patch}} 
\approx R_\odot^2 \frac{\int \xi \delta F(\xi) \rho dr}{M_{\rm patch}}
\; ,
\label{eq.pert_theory}
\end{equation}
where
\begin{equation}
M_{\rm patch}=\int \xi^2 \rho r^2 dr \; 
\label{eq.modemass}
\end{equation}
is the mode mass of the patched (solar) model, which is defined later in this subsection,
and where we have assumed that the perturbation is confined close to the surface, allowing us to replace $r^2$ with $R_\odot^2$ in the integral in the numerator of Eq. \ref{eq.pert_theory}.
For the extended model, we similarly get
\begin{equation}
\delta\omega_{\rm extend}^2 = \frac{\int \xi \delta F(\xi) \rho dz}{M_{\rm extend}} \; .
\label{eq.pert_extend}
\end{equation}
As the perturbations we consider (analogous to $\delta F$) are confined very close to the surface and well within the extended model, the numerators in Eqs. \ref{eq.pert_theory} and \ref{eq.pert_extend} should be identical for the Sun and for the \muram case. 
From this, it follows that the frequency perturbation in the solar case is given by:
\begin{equation}
\label{eq:scale}
\delta\omega  = \frac{R_\odot^2 M_{\rm extend}}{M_{\rm patch}} \delta\omega_{\rm extend} \; ,
\end{equation}
where we have assumed the approximation $\delta\omega^2 \approx 2\omega\delta\omega$, which is valid for $\delta\omega \ll \omega$.
The mode masses $M_{\rm extend}$ and $M_{\rm patch}$ have different units.  This is because one is a mode-mass for a plane-parallel model (\Mextend) and the other is for a spherical model.

It is essential that the eigenfunctions are identically normalized in the two cases
in order to use the scaling of equation~(\ref{eq:scale}). 
To achieve this,
we first patch the outermost part of the averaged MURaM stratification onto Model~S \citep{1996Sci...272.1286C}.
This is done by shifting the averaged simulation in height, so as to, on average, match the pressure and density at a depth of 1.5~Mm in Model~S. Model~S and the averaged simulation are then averaged between depths of 1~Mm and 2~Mm by a weight that varies linearly from 100\% Model S at 2~Mm to 100\% simulation average at 1~Mm, where MURaM and Model S match well. Using a gradual transition, as opposed to a step change, smooths out any remaining discontinuities.  We denote this model \Mpatch.  Based on this, we then solve the oscillation equations again (eq.~\ref{eq.dxidz} and~\ref{eq.dpdz}), together with the boundary conditions of regularity at the center of the star and the same top boundary condition that is used for the Cartesian box in order to obtain consistent mode masses. For each resonant mode, we use the eigenfunction that results from this calculation to compute the mode mass $M_{\rm patch}$ for the the model \Mpatch. We then calculate the scaled frequency perturbation using equation~(\ref{eq:scale}), having interpolated the mode masses $M_{\rm patch}$ to the frequency grid of the $\delta\omega_{\rm extend}$.

An important effect of the mode-mass scaling of the simulation results is that it eliminates the sensitivity linked to how the model is extended below the simulation domain and that the results are independent of which zero crossing is selected, as demonstrated numerically in Appendix \ref{app:tests}. We also emphasize that the only use of the patched model is to calculate the mode masses, which are used in the scaling. As such, small errors are of little consequence.

\section{Results}
\label{sec:results}

Figure~\ref{fig:solar} shows the frequency perturbations computed using equation~(\ref{eq:scale}) for the various simulations.  
These frequency perturbations are due to all of the physics not captured in the patched model \Mpatch.
As cases~10, 12, and 13 have the same resolution, the results have been connected for clarity.
The error bars were estimated from equation~(2) of \cite{1992ApJ...387..712L}
using the estimated linewidths, zero background noise, and then scaled based on the mode mass as was done for the frequency perturbations.
They do not include the contribution from the noise in the measured eigenfunctions.
However, the smooth variation of the frequency shift between simulations with different depths indicate that other effects do not contribute significant noise relative to the observed effect.
More importantly, they do not include any systematic contributions, such as those from the imperfect match between the observed and the analytical eigenfunctions.

As noted earlier, we see that modes with other $k_h$ leak weakly into $k_h=0$.
To ensure that these do not bias the results, we inspected the spectra of $k_h=1$ and $k_h = \sqrt{2}$ (with $k_{\rm h}$ in units of wavelengths across the box width) to determine if any peaks fall within the fitting interval. As it turns out, none are close to the fitting window below 3400~$\mu$Hz. Above this frequency, some of the leaks are within fitting window. This is largely due to the rapidly increasing linewidth. We therefore do not trust the results above 3400~$\mu$Hz, as indicated in Fig. \ref{fig:solar}

As part of the procedure, a number of choices were made, including which zero crossing to use as the lower boundary condition, the depth from which to extend the model analytically (given by $z_{\rm extend}$), $\Gamma_1$ in the extension, and the fitting interval (given by $z_{\rm fit,min}$ and $z_{\rm fit,max}$). While these would not be expected to have a large effect if chosen reasonably, we nonetheless study the effect of changing various parameters in Appendix \ref{app:tests}.
As expected, most parameters have a negligible effect, except for the fitting range where, not surprisingly, small changes are seen. However, these are small relative to the remaining discrepancies when not close to the photosphere, so they are of little importance for the present study.

The results from case~11 agree well with those of case~10, while those of simulation 15, which has poorer resolution, show significant differences. 
In other words, it appears that the resolution of case~10 (and thus~12 and 13) is adequate for the purpose of this study. 

\subsection{Comparison with previously published residuals}

\begin{figure}
\begin{center}

\begin{tikzpicture}[x=0.11\columnwidth,y=0.11\columnwidth]
\node at (0, 0) {\includegraphics[width=0.95\columnwidth]{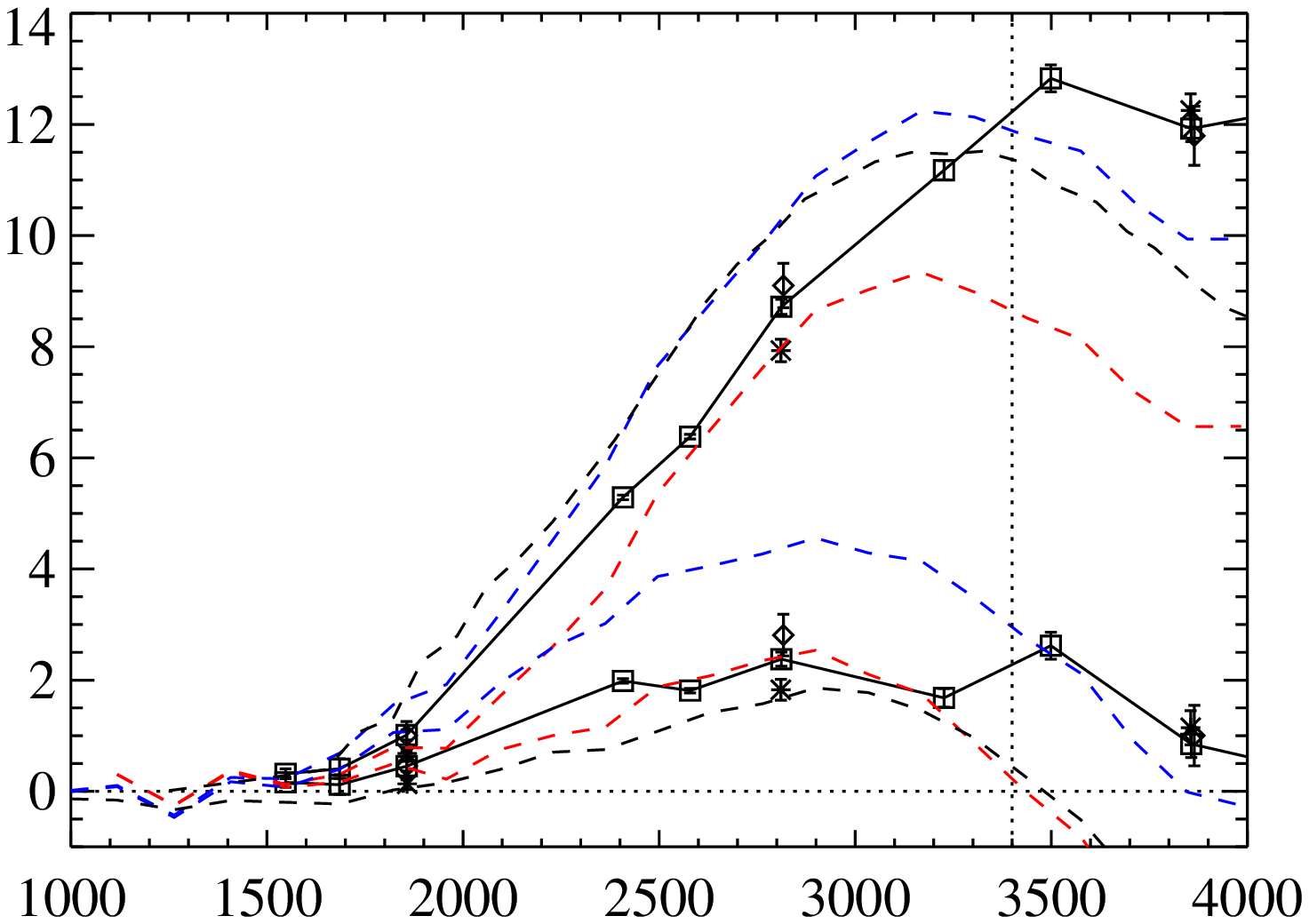}};
\node at (1, -3) {Frequency ($\mu$Hz)};
\node[rotate=90] at (-3.7,0.3) {$\delta\nu$ ($\mu$Hz)};
\end{tikzpicture}
\end{center}\caption[]{
Frequency perturbations scaled to the Sun, plotted as a function of the simulation frequencies.
Squares are for cases 10, 12, and 13, diamonds are for case~11, and crosses are for case~15.
The solid lines connect the result of cases~10, 12, and 13. The top curve shows the RGM cases; the bottom curve shows the GGM results.
The results for the first zero crossing below 10~Mm were used (as opposed to a shallower crossing) in order to ensure that the physics is well represented by the classic equations. The effects of using different crossings are discussed in Appendix \ref{app:tests}.
For the sake of comparison, the upper dashed black line shows the residuals (Sun minus model)
from the RGM patched model of \citet[model "B"]{2017MNRAS.464L.124H} 
and the lower dashed black line shows the residuals from the GGM patched model of \citet{2016A&A...592A.159B}.
Regarding the latter, we note that we were unable to reproduce the frequencies from the model (obtained from Ball, private communications) exactly, due to what appears to be numerical problems in the model. However, the differences were small enough so as to not affect the overall conclusion.
The red dashed lines show the results from Fig. 5 of \cite{2018MNRAS.481L..35J}, while the blue lines show the RGM numbers shown in blue in Fig. 6 of \cite{2019MNRAS.488.3463J}. The corresponding GGM numbers are courtesy of J{\o}rgensen (private communication).
The numbers from \cite{2017A&A...600A..31S} are not shown as they have deviations at low frequencies, which are indicative of an overall structural error.
The vertical dotted line indicates roughly where leaks from other $k_h$ start to appear within the fitting interval, as is discussed in the main text.
\label{fig:solar}}
\end{figure}

Figure~\ref{fig:solar} also shows the frequency residuals (Sun minus model) for the patched model with turbulent pressure and the RGM approximation from \citet[model "B," dashed line in their Fig.~1]{2017MNRAS.464L.124H}.
In the middle of the p-mode band (around 3000~\muHz), we explain almost all of their residuals with our RGM calculation.  
At lower frequencies, we clearly overestimate the residuals, and  we underestimate them at high frequencies. Given the results in the next subsection, we did not compare our model with the more elaborate model used by \cite{2017MNRAS.464L.124H} directly.
For a comparison, the RGM numbers of \cite{2018MNRAS.481L..35J} are somewhat smaller, while the numbers from \cite{2019MNRAS.488.3463J} are a bit larger than those of \cite{2017MNRAS.464L.124H}.

As another test of the model presented here,
Fig.~\ref{fig:solar} shows the residuals from the GGM patched model
from \citet[Fig. 1, open circles]{2016A&A...592A.159B}.
Around 3000~\muHz, we explain almost all of these residuals. At low frequencies, where the residuals are small and the errors are relatively large, there is no clear improvement. 
At frequencies above about 3400~\muHz, our results provide a poor match; however, as noted earlier, this may be due to leaks from other modes.
For a comparison, the GGM numbers of \cite{2018MNRAS.481L..35J} are somewhat smaller, while the GGM numbers of \cite{2019MNRAS.488.3463J} are larger than the numbers from \cite{2016A&A...592A.159B}.

\subsection{Comparison with Houdek et al. (2017) eigenfunctions}
\label{sec:houdek-comp}

\begin{figure}
\begin{center}

\begin{tikzpicture}[x=0.11\columnwidth,y=0.11\columnwidth]
\node at (0, 0) {\includegraphics[width=0.95\columnwidth]{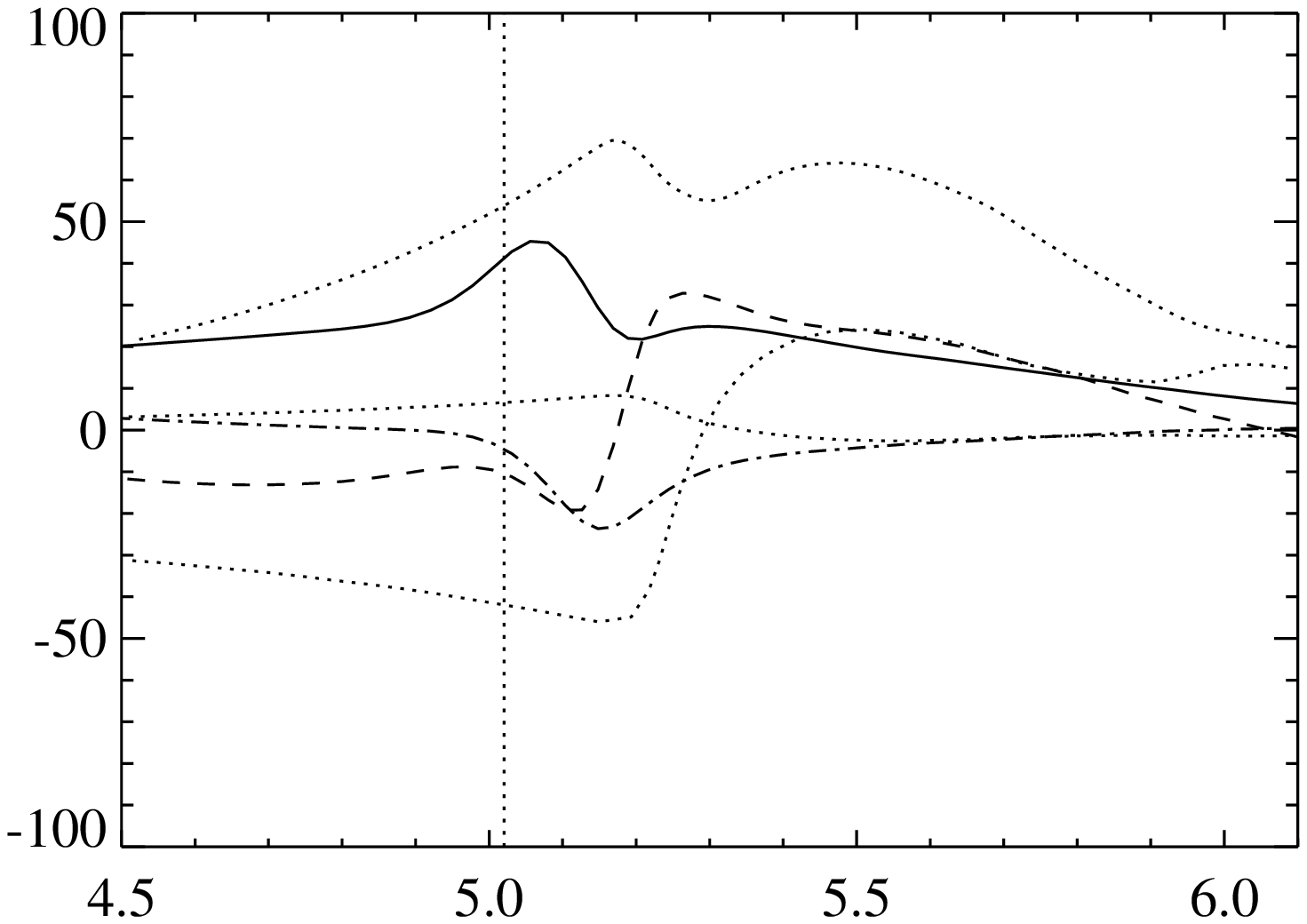}};
\node at (1, -3) {$\log_{10} p_{\rm tot}$};
\node[rotate=90] at (-3.7,0.3) {phase (deg), $|\delta\pturb/\overline{p}_{\rm tot}|$};
\end{tikzpicture}
\end{center}\caption[]{
Quantities from Fig.~2 of \cite{2017MNRAS.464L.124H} for the mode at 2815~\muHz from case~11.
The dashed line shows the phase difference between the Lagrangian density perturbation and the Lagrangian turbulent pressure perturbation.  The
dash-dotted line shows the phase difference relative to the Lagrangian gas pressure perturbation.
The solid line shows the norm of the ratio of the Lagrangian turbulent pressure perturbation to the total pressure, which was multiplied by an arbitrary factor to match the range of the other quantities.  The dotted lines show the same three quantities from \cite{2017MNRAS.464L.124H}.
\label{fig:houdek2}}
\end{figure}

As discussed earlier, we can study the approximations suggested by \cite{2017MNRAS.464L.124H} since we have access to the eigenfunctions of the relevant quantities. Figure~\ref{fig:houdek2} shows the quantities displayed in Fig.~2 of \cite{2017MNRAS.464L.124H} for a mode with a frequency that is close to the mode used for that figure; the frequency is quite close and the quantities do not strongly depend on frequency. 
The correspondence is rather poor.
The phases of the pressure perturbation eigenfunctions agree quite well over the log pressure range between 5.5 and 5.9, but they are very different closer to the surface.
Also, the norm of the pressure perturbation eigenfunction has a substantially different behavior from what \cite{2017MNRAS.464L.124H} predict.
We note that the curves agree better if they are shifted horizontally (i.e., shifted in depth). This may be due to the fact that the background state used is somewhat inaccurate (G. Houdek, private communications).

Given this situation, it is perhaps somewhat surprising that the frequency perturbations calculated by \cite{2017MNRAS.464L.124H} are quite good. The discrepancy suggests that the
results of applying the theory to stars of a different spectral type should be used with caution.

\section{Discussion}
\label{sec:disc}

As we have demonstrated, our approach allowed us to determine the surface term quite well. 
In particular, we were able to estimate the surface term without
relying on the accuracy of the GGM or RGM approximations, both of which lack a good physical justification.

Having said that, the fits are not perfect, and this is particularly the case at high frequencies. There are several potential reasons for this. While the errors on the observed solar frequencies are negligible relative to the residuals over most of the frequency range, the effects of line asymmetry may provide some contribution.
The residuals shown in Fig.~\ref{fig:solar} are based on the differences to a solar model and may thus contain model errors. While they are difficult to assess, these errors are also unlikely to explain the discrepancies.

While we implemented GGM and RGM corrections, as used by \cite{2016A&A...592A.159B}, \cite{2017MNRAS.464L.124H}, and others, and while \cite{2016A&A...592A.159B} essentially used the same MURaM models as those employed for the purposes of this study, we did not use the exact same models and we did not implement the exact same boundary conditions.
While these differences are expected to be small, they are both located very close to the surface and the resulting frequency changes will thus likely appear as a surface term. Given this they may explain all or much of the discrepancies, especially at high frequencies.

Finally, it is possible that either our models or those we compare to have inaccuracies in the physics. Possible problems include incorrect opacities, equation of state, or approximations in the radiative transfer.

While our results are quite promising and show that the suggested approach to modeling the surface term works, there are many possibilities for improvements, ranging from technical ones to a better understanding of the underlying physics.
On the technical side, the lack of a dense grid of frequencies is clearly  a problem. Running simulations with different depth ranges, as done here, is one way to mitigate this problem. Unfortunately, this is very computationally expensive. More fundamentally, it involves extending the simulations to depths where the physics is simple and well described by the classic equations and is thus wasteful.

An alternative approach would be to change the bottom boundary condition or to put a partly reflecting layer inside the simulation. Both of these options would shift the resonance frequencies of the computational domain. These solutions would, of course, need to be carried out with extreme care in order to not affect the background state by applying a change that only affects the dynamics at high frequencies, for example.
A related technical problem is the long simulation times needed
to reduce the effect of the stochastic excitation and rather large damping on our ability to measure the frequencies.

An intriguing way around these problems would be to add a source of excitation of the acoustic waves well below the part of the simulation with nonstandard physics. In this way, one or more well defined frequencies could be excited, possibly at a large amplitude and covering the desired range of frequencies. This approach would
eliminate the problem of estimating the frequencies.
Unfortunately, simply driving at a frequency that is far from a resonance is, at best, problematic; the presence of nodes in the eigenfunctions of the perturbed quantity requires careful placement of the excitation and the simulations still have naturally excited modes.
A way to circumvent these problems would be to add a damping layer near the bottom of the simulations. Even if imperfect, this would allow for driving at any desired frequency.
Of course, the concerns regarding a possible modification of the background need to be kept in mind.
Also, it is important to realize that the stochastic nature of the convection still affects the interaction of the modes with the convection.

Another thing to study would be nonradial ($k_h \neq 0$) modes. However, for the global modes used for inversions of solar-like oscillations in Sun-like stars, the results are not expected to be significantly different.

While we have chosen to divide the procedure into two parts, one to estimate the frequency perturbation in a shallow box and one to scale it to the Sun, it may be possible to combine these by actually patching the eigenfunctions or by deriving an effective boundary condition at some suitable point, for instance.
While this may seem attractive and it avoids the mixing of the patching approach and the perturbation approach, there are significant complications.
If the matching point is chosen close to the surface, the classic equations do not work well there.
If chosen deeper down, curvature and variable gravity effects have to be considered.
The approach we have taken carefully avoids these issues, at the price of, perhaps, being conceptually more complicated.

Toward the more theoretical side, studying the numerical eigenfunctions may allow us to obtain significant insights, as already shown in Sect.~\ref{sec:houdek-comp}.  A physical understanding of the wave-convection interaction, perhaps along the lines of \cite{2017MNRAS.464L.124H}, would clearly be highly desirable. However, this would also probably  be difficult.
The approach of \citet{2013ApJ...773..101H} is another promising possibility.

A better understanding of the interactions is also important for other problems in helio- and asteroseismology, such as the center-to-limb effects present in helioseismology and the visibilities of oscillation modes in other stars \citep{2018A&A...617A.111S}. Indeed the former was the original motivation for \cite{2012ApJ...760L...1B}.
Having said that, the most important application is likely to be in asteroseismology, where the need to fit for the surface term or make questionable assumptions is currently impeding progress.

\section{Conclusion}
\label{sec:conc}

We have shown that it is possible to estimate the surface term to within about 1~\muHz in the middle of the p-mode band using hydrodynamic simulations.

However, it is also clear that further work is needed before the results can be applied in helio- and asteroseismic inversions. In particular, the remaining discrepancies need to be understood, a way needs to be found to obtain a denser frequency grid, and computations have to be made for a range of stars.

\begin{acknowledgements}
The authors would like to thank Robert Cameron for assistance with running the MURaM simulations and for general discussions, Warrick Ball for help with understanding details of \cite{2016A&A...592A.159B} and for providing values from plots, G\"unter Houdek for explaining details of \cite{2017MNRAS.464L.124H} and various physics issues,
and Regner Trampedach for many useful discussions.
The authors also thank Andreas Christ S{\o}lvsten J{\o}rgensen and Jakob Mosumgaard  for providing us with the data shown in Fig. 4 and for many useful discussions.
We acknowledge partial support from the European Research Council Synergy Grant WHOLE SUN $\#$810218. 
\end{acknowledgements}

%-------------------------------------------------------------------

\bibliographystyle{aa}
\bibliography{surf19}

\begin{appendix}
\section{Tests of various approximations}
\label{app:tests}

In the main part of this paper, we have made a number of approximations and somewhat arbitrary choices (such as the matching depth). While the approximations would be expected to be reasonable and the overall approach should be insensitive to the choices, it is nonetheless instructive to examine some of them in detail, and we do so in the following subsections.
We note that the smooth variation of the scaled perturbations with the simulation depth already validates some of the choices, and we have already discussed the effect of the spatial resolution.
For the tests, we used the GGM version of case~12 as this has more frequencies and can be truncated to correspond to cases~10 and 13.

\subsection{Zero crossing choice}
Table \ref{table:zero} shows the results of using different zero crossings. As expected, the unscaled frequency changes vary substantially, but this is compensated by the change in the mode mass, such that the resulting differences are negligible, even when using crossings that are very close to the surface.

\begin{table}
\caption{
Unscaled and scaled frequency changes resulting from different choices for which zero crossing, $z_{\rm zero}$, was used to define the lower boundary condition.
Boldfaced values are the standard ones used in the main text.
Other parameters were kept at their standard values for case~12.
}
\label{table:zero}      
\centering                          
\begin{tabular}{|c r c c r c c|}        
\hline                 
n & $z_{\rm zero}$ & $\nu_{\rm sim}$ & $\nu_{\rm extend}$ & $\delta\nu_{\rm extend}$ & \underline{$R_\odot^2 M_{\rm extend}$} & $\delta\nu$ \\
 & Mm & \muHz & \muHz & \muHz & $M_{\rm patch}$ & \muHz \\
\hline                        
0 & -27.3 & 1547.54 & 1546.97 &  0.57 & 0.281 & 0.16 \\
0 & {\bf -10.3} & 1547.54 & 1546.65 &  0.89 & 0.180 & 0.16 \\
\hline
1 & -46.1 & 2407.75 & 2402.32 &  5.44 & 0.363 & 1.97 \\
1 & -30.1 & 2407.75 & 2401.33 &  6.42 & 0.308 & 1.98 \\
1 & {\bf -17.6} & 2407.75 & 2399.89 &  7.86 & 0.252 & 1.99 \\
1 &  -8.7 & 2407.75 & 2397.55 & 10.21 & 0.196 & 2.00 \\
1 &  -3.2 & 2407.75 & 2392.23 & 15.52 & 0.130 & 2.02 \\
\hline
2 & -43.7 & 3226.47 & 3221.58 &  4.89 & 0.343 & 1.68 \\
2 & -31.8 & 3226.47 & 3220.89 &  5.57 & 0.301 & 1.68 \\
2 & -21.9 & 3226.47 & 3219.98 &  6.48 & 0.259 & 1.68 \\
2 & {\bf -14.0} & 3226.47 & 3218.71 &  7.75 & 0.217 & 1.68 \\
2 &  -8.0 & 3226.47 & 3216.79 &  9.67 & 0.174 & 1.68 \\
2 &  -4.0 & 3226.47 & 3213.28 & 13.18 & 0.128 & 1.68 \\
2 &  -1.3 & 3226.47 & 3204.07 & 22.40 & 0.075 & 1.69 \\
\hline                                   
\end{tabular}
\end{table}

\subsection{Model extrapolation}
Another somewhat arbitrary choice is how to extrapolate the models downward. 
Based on the discussion in Sec. \ref{sec:sun}, the deeper stratification should not affect the final results as the resulting changes in the resonant frequencies would be compensated for by the corresponding changes in mode masses.
The most obvious choice we made was how to extrapolate downward. In the main part of the paper, we chose to linearly increase $\Gamma_1$ with depth to roughly match deeper simulations. We then assumed adiabatic stratification, constant gravity, and hydrostatic equilibrium to obtain the structure.
As an alternative, we kept $\Gamma_1$ constant, resulting in a large change in the sound speed. The results are given in Table \ref{table:gamma1}. While the positions of the zero crossings change significantly, the resulting frequency changes are negligible, giving us further confidence in the insensitivity to the deeper structure.

\begin{table}
\caption{
Unscaled and scaled frequency changes for the standard $\Gamma_1$ extrapolation (top half) and constant $\Gamma_1$ (bottom half).
Other parameters were kept at their standard values for case~12.
The same crossings were used in each case, but for $n=0$ it moved just inside the normal 10~Mm limit due to the change in stratification.
}
\label{table:gamma1}      
\centering                          
\begin{tabular}{|c r c c r c c|}        
\hline                 
n & $z_{\rm zero}$ & $\nu_{\rm sim}$ & $\nu_{\rm extend}$ & $\delta\nu_{\rm extend}$ & \underline{$R_\odot^2 M_{\rm extend}$} & $\delta\nu$ \\
 & Mm & \muHz & \muHz & \muHz & $M_{\rm patch}$ & \muHz \\
\hline               
0 & -10.3 & 1547.54 & 1546.65 &  0.89 & 0.180 & 0.16 \\
1 & -17.6 & 2407.75 & 2399.89 &  7.86 & 0.252 & 1.99 \\
2 & -14.0 & 3226.47 & 3218.71 &  7.75 & 0.217 & 1.68 \\
\hline
0 &  -9.8 & 1547.54 & 1546.64 &  0.90 & 0.178 & 0.16 \\
1 & -16.2 & 2407.75 & 2399.94 &  7.82 & 0.254 & 1.99 \\
2 & -13.2 & 3226.47 & 3218.73 &  7.73 & 0.217 & 1.68 \\
\hline                                   
\end{tabular}
\end{table}

\subsection{Fitting interval}
As shown in Fig. \ref{fig:efit}, the eigenfunctions from the simulation and the analytical ones do not match perfectly. Given this, some effect of the fitting range may be expected; therefore, we explore the effect in 
Table \ref{table:fit}.
As expected, there is some variation, especially when fitting close to the surface, but over the range used for the different cases, the changes are small relative to the remaining discrepancies.

\begin{table}
\caption{
Unscaled and scaled frequency changes for different choices of where to match the simulation and analytical eigenfunctions.
In all cases $z_{\rm fit,max}$ is 1~Mm above $z_{\rm fit,min}$.
Other parameters were kept at their standard values for case~12.
We note that the fits for the shallowest two cases extend across or are close to $z_{\rm extend} = -5.5$~Mm.
}
\label{table:fit}      
\centering                          
\begin{tabular}{|c c c c r r|}        
\hline                 
n & $z_{\rm fit,min}$ & $\nu_{\rm sim}$ & $\nu_{\rm extend}$ & $\delta\nu_{\rm extend}$ & \multicolumn{1}{c}{$\delta\nu$} \\
 & Mm & \muHz & \muHz & \muHz  & \muHz \\
\hline                        
0 &  -6.0 &  1547.54 &  1546.90 &   0.64 &   0.11 \\
0 &  -5.5 &  1547.54 &  1546.92 &   0.62 &   0.11 \\
0 &  {\bf -5.0} &  1547.54 &  1546.65 &   0.89 &   0.16 \\
0 &  -4.5 &  1547.54 &  1546.38 &   1.16 &   0.21 \\
0 &  -4.0 &  1547.54 &  1546.04 &   1.50 &   0.27 \\
0 &  -3.5 &  1547.54 &  1545.14 &   2.40 &   0.44 \\
0 &  -3.0 &  1547.54 &  1547.43 &   0.11 &   0.02 \\
0 &  -2.5 &  1547.54 &  1547.04 &   0.50 &   0.09 \\
0 &  -2.0 &  1547.54 &  1546.85 &   0.69 &   0.12 \\
0 &  -1.5 &  1547.54 &  1546.73 &   0.81 &   0.15 \\
0 &  -1.0 &  1547.54 &  1547.44 &   0.10 &   0.02 \\
\hline
1 &  -6.0 &  2407.75 &  2399.86 &   7.90 &   1.99 \\
1 &  -5.5 &  2407.75 &  2399.83 &   7.93 &   2.00 \\
1 &  {\bf -5.0} &  2407.75 &  2399.89 &   7.86 &   1.99 \\
1 &  -4.5 &  2407.75 &  2399.90 &   7.86 &   1.98 \\
1 &  -4.0 &  2407.75 &  2399.92 &   7.83 &   1.98 \\
1 &  -3.5 &  2407.75 &  2399.92 &   7.83 &   1.98 \\
1 &  -3.0 &  2407.75 &  2399.99 &   7.76 &   1.96 \\
1 &  -2.5 &  2407.75 &  2400.58 &   7.18 &   1.81 \\
1 &  -2.0 &  2407.75 &  2400.81 &   6.95 &   1.75 \\
1 &  -1.5 &  2407.75 &  2397.95 &   9.81 &   2.49 \\
1 &  -1.0 &  2407.75 &  2400.39 &   7.37 &   1.86 \\
\hline
2 &  -6.0 &  3226.47 &  3218.86 &   7.61 &   1.65 \\
2 &  -5.5 &  3226.47 &  3218.79 &   7.68 &   1.66 \\
2 &  {\bf -5.0} &  3226.47 &  3218.71 &   7.75 &   1.68 \\
2 &  -4.5 &  3226.47 &  3218.67 &   7.80 &   1.69 \\
2 &  -4.0 &  3226.47 &  3218.77 &   7.69 &   1.67 \\
2 &  -3.5 &  3226.47 &  3219.08 &   7.39 &   1.60 \\
2 &  -3.0 &  3226.47 &  3219.43 &   7.04 &   1.52 \\
2 &  -2.5 &  3226.47 &  3219.26 &   7.20 &   1.56 \\
2 &  -2.0 &  3226.47 &  3218.10 &   8.36 &   1.81 \\
2 &  -1.5 &  3226.47 &  3217.91 &   8.56 &   1.85 \\
2 &  -1.0 &  3226.47 &  3222.12 &   4.35 &   0.94 \\
\hline                                   
\end{tabular}
\end{table}

\end{appendix}

\end{document}